\documentclass{emulateapj}

\def\Journal#1#2#3#4{{#4}, {#1}, {#2}, #3}

\def\AAA{A\&A}
\def\ApJ{ApJ}

\def\Aph{Astropart. Phys.}
\def\ApJS{ApJSS}
\def\ML{Machine Learning}
\def\MNRAS{MNRAS}

\def\NIMA{Nucl. Instrum. Methods A}

\def\SCI{Science}

\shorttitle{1ES~1011+496}
\shortauthors{Albert et al.}

\begin{document}


\title{Discovery of Very High Energy $\gamma$-rays from 1ES~1011+496 at $z=0.212$}


\author{J.~Albert\altaffilmark{a}, 
 E.~Aliu\altaffilmark{b}, 
 H.~Anderhub\altaffilmark{c}, 
 P.~Antoranz\altaffilmark{d}, 
 A.~Armada\altaffilmark{b}, 
 C.~Baixeras\altaffilmark{e}, 
 J.~A.~Barrio\altaffilmark{d},
 H.~Bartko\altaffilmark{f}, 
 D.~Bastieri\altaffilmark{g}, 
 J.~K.~Becker\altaffilmark{h}
% ...
}
 \altaffiltext{a} {Universit\"at W\"urzburg, D-97074 W\"urzburg, Germany}
 \altaffiltext{b} {IFAE, Edifici Cn., E-08193 Bellaterra (Barcelona), Spain}
 \altaffiltext{c} {ETH Zurich, CH-8093 Switzerland}
 \altaffiltext{d} {Universidad Complutense, E-28040 Madrid, Spain}
 \altaffiltext{e} {Universitat Aut\`onoma de Barcelona, E-08193 Bellaterra, Spain}
 \altaffiltext{f} {Max-Planck-Institut f\"ur Physik, D-80805 M\"unchen, Germany}
 \altaffiltext{g} {Universit\`a di Padova and INFN, I-35131 Padova, Italy}  
 \altaffiltext{h} {Universit\"at Dortmund, D-44227 Dortmund, Germany}
% ...

\begin{abstract}
We report on the discovery of Very High Energy (VHE) $\gamma$-ray
emission from the BL Lacertae object 1ES~1011+496. The observation was
triggered by an optical outburst in March 2007 and the source was
observed with the MAGIC telescope from March to May 2007. 
Observing for 18.7 hr, we find an excess of
6.2\,$\sigma$ with an integrated
flux above 200\,GeV of (1.58$\pm0.32)\cdot 10^{-11}$ photons cm$^{-2}$
s$^{-1}$. 
The VHE $\gamma$-ray flux is $>40\%$ higher than in March--April 2006
(reported elsewhere), indicating that the VHE emission state may be
related to the optical emission state. We have also determined the
redshift of 1ES~1011+496 based on an optical spectrum that reveals the
absorption lines of the host galaxy. The redshift of $z=0.212$ makes
1ES~1011+496 the most distant source observed to emit VHE
$\gamma$-rays to date.
\end{abstract}

\keywords{gamma rays: observations, quasars: individual (1ES~1011+496)}

\section{Introduction}
Known Very High Energy (VHE defined as $>100$GeV) $\gamma$-ray
emitting Active Galactic Nuclei (AGN) show variable flux in all
wave bands. The relationship between the variability in different
wave bands appears rather complicated. The MAGIC collaboration is performing
Target of Opportunity (ToO) observations whenever alerted that sources
are in a high flux state in the optical and/or X-ray bands. 
Previously, optically triggered observations resulted in the
discovery of VHE $\gamma$-ray emission from Markarian~180
\citep{mrk180}. Here we report the discovery of VHE $\gamma$-rays from
1ES~1011+496 triggered by an optical outburst in March 2007.
Previous observations of the source with the MAGIC telescope did not 
show a clear signal \citep{hbl}.

1ES~1011+496 is a high frequency peaked BL~Lac (HBL) object for which
we now determined a redshift of $z=0.212\pm0.002$
(Fig.~1). Previously, this was uncertain since it was based on an
assumed association with the cluster Abell 950 \citep{wisniewski}. The
redshift determined here makes 1ES~1011+496 the most distant 
VHE $\gamma$-ray source yet detected 
with the possible exception of PG~1553+113
\citep{hess1553, magic1553} (for which the redshift 
is $0.09<z<0.42$ \citep{sba,mazgoe}).


\begin{figure}
\includegraphics[width=0.45\textwidth]{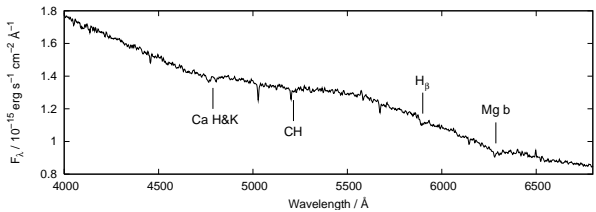}
\caption{Optical spectrum of 1ES~1011+496 obtained with the Multi
  Mirror Telescope, using the Blue Channel Spectrograph with the 300
  line/mm grating, a 1.5'' slit, and Loral 3kx1k CCD. Integration time
  was 30 minutes. Absorption lines of the host galaxy (Ca H\&K, CH G, H$_{\beta}$ and Mg b) are clearly visible and indicate a redshift of $z=0.212\pm0.002$.}
\end{figure}


The spectral energy distribution (SED) of BL~Lac objects normally shows a
two-bump structure. The lower frequency peak is due to 
synchrotron radiation. Various models have been proposed for the
origin of the high-frequency peak; the most popular invoke
inverse Compton scattering of ambient soft photons.
There have been several suggestions for
the origin of the low-frequency seed photons that are up scattered
to $\gamma$-ray energies: the soft photons may be produced within the jet
itself by synchrotron radiation (SSC, \citet{maraschi}) or come from
outside the jet, perhaps from the accretion disk (EC, \citet{dermer}). The
high-energy peak may, instead, also have a hadronic origin (e.g. \citet{mannheim}).

When the synchrotron emission peak is located in the low energy range from the 
sub-millimeter to optical, the objects are called low-frequency-peaked BL~Lac objects. HBLs,
on the other hand, have the peak synchrotron emission in the UV to
X-ray energy range. The peak of the second bump is often not
observable because of the low sensitivity above a few hundred MeV of
satellite-borne detectors or a too high energy threshold of
ground-based $\gamma$-ray detectors. With the exception of M\,87
\citep{hegram87,hessm87} and BL~Lac \citep{bllac}, all
known blazar sources detected at TeV energies with Cherenkov
telescopes show a synchrotron peak in the UV to X-ray energy range,
suggesting that the intensity of the TeV emission is related to a
synchrotron component extending to high frequencies.

\section{Observations and Data Analysis}
The MAGIC telescope is located on the Canary Island La Palma
(2200 m above sea level, 28$^\circ$45$'$N, 17$^\circ$54$'$W).  
The accessible energy range spans from~50-60 GeV 
(trigger threshold at small zenith angles) up to tens of TeV \citep{crab}.

The MAGIC observation was triggered by an observed high optical state
of 1ES~1011+496 on 2007 March 12 (see light curve Fig.~2). The source
has been monitored for more than 4 years in the optical
with the KVA\footnote{http://tur3.tur.iac.es} and Tuorla 1\,m telescopes as a
part of the Tuorla blazar monitoring
program.\footnote{http://users.utu.fi/$\sim$kani/1m/} In March 2007 the flux
reached the highest level ever observed during the
monitoring. The core flux, which is the host-galaxy-subtracted flux
(the host galaxy flux is taken from \citet{nilsson07} and is
$0.49\pm0.02$~mJy), increased more than $50\%$ from the local minimum
of the light curve. The high optical state with increasing flux was
continuing throughout the MAGIC observations, despite an observation gap
of 3 weeks due to bad weather.

1ES~1011+496 is monitored by \textit{RXTE} ASM and \textit{Swift} BAT, but the X-ray
flux of the source is below the sensitivity of these
instruments and the light curves show no indication of flaring. 
The source was also observed at Mets\"ahovi Radio Observatory in May 2007.
The source was not detected at 37~GHz, which indicates that it was not
in a high state at millimeter wavelengths (A. L\"ahteenm\"aki 2007,
private communication).

\begin{figure}
\includegraphics[width=0.15\textwidth, angle=270]{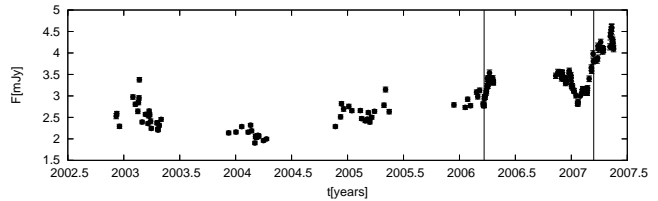}
\caption{The optical R-band light curve of 1ES~1011+496 from Tuorla 1 meter and KVA 35cm telescopes. The vertical lines indicate the starting point of the MAGIC observations in 2006 and 2007.}
\end{figure}


After the alert, MAGIC observed 1ES~1011+496 in March--May 2007.  
The total observation time was 26.2 hr, and the observation was
performed at zenith angles ranging from $20\degr$ to
$37\degr$. 
The observation was done in the so-called Wobble-mode
\citep{daum}. After removing runs with unusual trigger rates, mostly
caused by bad weather conditions, the effective observational time
amounts to 18.7 hr.


The data were analyzed using the standard analysis and calibration
programs for the MAGIC telescope \citep{crab}.  
The analysis is based on image parameters \citep{hillas}, the Random
Forest \citep{breiman,rf}, and the DISP methods \citep{domingo}.
After cuts for $\gamma$/hadron separation, the distribution of the
angle $\theta$, which is the angular distance between the source
position in the sky and the reconstructed shower origin, is used to
determine the signal in the ON-source region. Three background (OFF)
regions of the same size are chosen symmetrically to the ON-source
region with respect to the camera center. The final cut
$\theta^2<0.02\,\deg^2$ to determine the significance (Fig.~3) was optimized on
nearly contemporaneous Crab data to determine the significance of the
signal and the number of excess events. 
The energy threshold was about 160\,GeV for this analysis, which, given the soft
spectrum of the source, allowed for signal extraction down to 100
GeV. 
The data were also analyzed with an independent analysis. Within the
statistical errors the same significance, flux, and differential
spectrum were obtained.

\section{Results}
The distribution of the $\theta^2$-values, after all cuts, is shown in
Fig.~3. The signal of 297 events over 1591 normalized background events
corresponds to an excess with significance of $6.2\sigma$ using
equation (17) in \citet{LiMa}.


\begin{figure}
\includegraphics[width=0.45\textwidth]{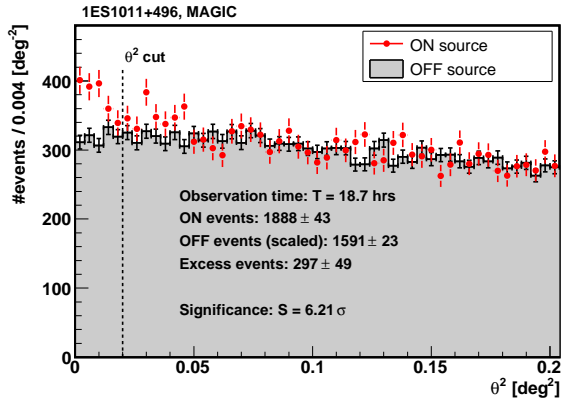}
\caption{Distribution of $\theta^2$ for ON-region data and normalized OFF-region data. The signal region is marked by the dashed line.}
\end{figure}


To search for time variability the sample was divided into 14
subsamples, one for each observing night. Fig.~4 shows the integral
flux for each night calculated for a photon flux above 200~GeV. The energy 
threshold has been chosen to reduce systematic effects arising from a 
rapidly decreasing effective area for $\gamma$-rays for lower energies. 
The flux is statistically constant at an emission level of
F($>$200~GeV)=(1.58$\pm0.32)\cdot 10^{-11}$ photons cm$^{-2}$s$^{-1}$. 

The energy spectrum of 1ES~1011+496 is shown in Fig.~5. It
extends from \mbox{$\sim 120$\,GeV} to \mbox{$\sim 750$\,GeV} and can
be well approximated by a power law:
\[
{\frac{\mathrm{d}N}{\mathrm{d}E}}=(2.0\pm0.1)\cdot10^{-10}\left({\frac{E}{0.2\,\mathrm{TeV}}}\right)^{-4.0\pm0.5}\,
\frac 1 {\mbox{TeV}\,\mbox{cm}^{2}\,\mbox{s}}  
\] 
The errors are statistical only. We estimate the systematic
uncertainity to be around 75\% for the absolute flux level
and 0.2 for the spectral index.
The observed spectrum is affected by the evolving extragalactic
background light (EBL, \citet{nikishov,stecker}) as the VHE
$\gamma$-rays are partially altered by interactions with the
low-energy photons of the EBL. Therefore, to obtain the intrinsic
spectrum of the source, the observed spectrum must be corrected.  The optical
depth and the resulting attenuation of the VHE $\gamma$-rays from
1ES~1011+496 are calculated using the number density of the evolving
EBL provided by the best-fit model of \citet{kneiske}. Within given 
model uncertainties, the model is in good agreement
with alternative models \citep{primack,steckerNEW} 
and EBL upper limits \citep{hessebl,mazinraue}.
Even after the correction, 
the slope of the spectrum is $\Gamma_{\mathrm{int}} = 3.3 \pm 0.7$ 
 (dashed brown line in Fig.~5,
$\chi^2\,/\,\mbox{NDF} = 2.55 / 2$), 
softer than observed for other HBLs in TeV energies and thus not providing new
constraints on the EBL density.


\begin{figure}
\includegraphics[width=0.45\textwidth]{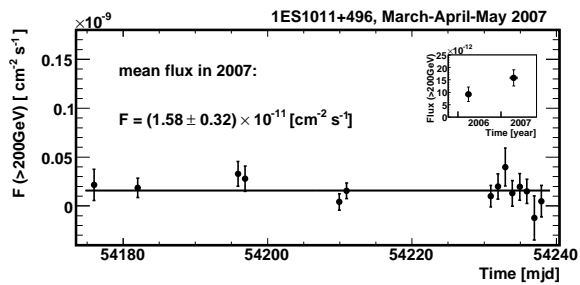}
\caption{The night-by-night light curve of 1ES~1011+496 from 2007 March 17 (MJD 54176) to 2007 May 18 (MJD 54238).
         The year-by-year light curve is shown in the inset, the 2006 data point is from \citet{hbl}.}
\end{figure}



\begin{figure}
\includegraphics[width=0.45\textwidth]{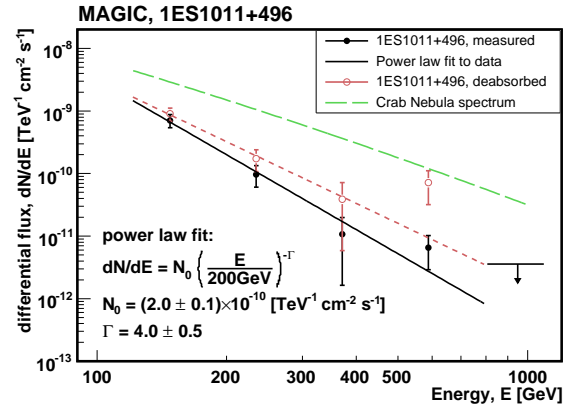}
\caption{The measured spectrum (black filled circles), the power-law
 fit to the data (solid line), the deabsorbed spectrum (brown
 open circles), and the fit to the deabsorbed spectrum
 (dashed brown line). The last measured point is a 95\% upper limit. 
 In the deabsorbed spectrum, the last spectral point at  $\approx$600\,GeV is 1.6\,$\sigma$
 above the fit and thus not significant. The Crab Nebula spectrum (green dashed line, \citet{crab}) 
 is shown for comparison.}
\end{figure}


\vspace{2cm}

\section{Discussion}
We report the discovery of VHE $\gamma$-ray emission from BL Lac
object 1ES~1011+496. With the redshift of $z=0.212$, it is the most
distant source detected to emit VHE $\gamma$-rays to date.  
The observed spectral properties 
(soft and no significant excess above $\sim 800$~GeV) 
are consistent with the state-of-the-art EBL models 
\citep{kneiske,primack,steckerNEW} and confirm recently derived 
EBL limits \citep{hessebl,mazinraue}. 

In Fig.~6 we show the SED of
1ES~1011+496 using historical data (open circles; 
\citet{costamante} and references therein) and our nearly simultaneous
optical \textit{R}-band data (triangle), together with the MAGIC spectrum
corrected for attenuation (filled
circles). We also display (square) the EGRET flux of the source
3EG~J1009+4855, which has been suggested to be associated 
with 1ES~1011+496 (\citet{hartman},
but see also \citet{soward} whose analysis disfavors the association).

We model the SED by using a one-zone synchrotron-SSC model (see
\citet{tavecchio01} for a description). In brief, a spherical emission
region is assumed with radius $R$, 
filled with a tangled
magnetic field of mean intensity $B$. The relativistic electrons follow a
broken-power-law energy distributions specified by the limits
$\gamma _{\rm min}$ and $\gamma _{\rm max}$ and the break at $\gamma
_{\rm b}$. Relativistic effects are taken into account by the Doppler
factor $\delta$.


\begin{figure}
\centering
\includegraphics[width=0.46\textwidth, bb = 40 200 590 670]{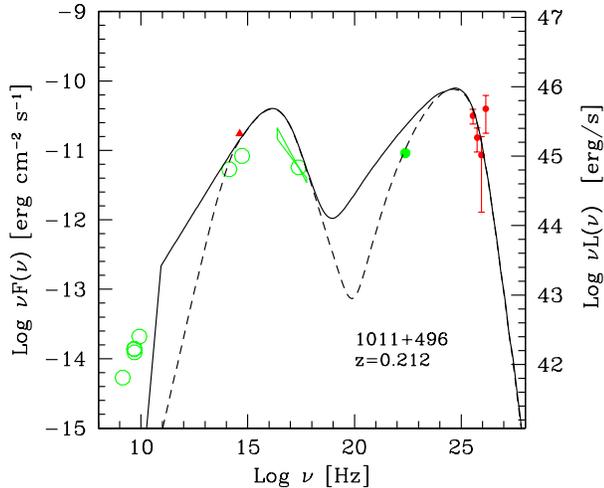}
\caption{Spectral energy distribution of 1ES~1011+496. The two
  different fits are done by varying the minimum electron energy 
  $\gamma_{\rm min}$ (see text). 
  The other fit parameters are: R(radius of sphere)=$10^{16}$\,cm,
  $\delta$(Doppler factor)=20, B(magnetic field) = 0.15\,G,  
   $\gamma_{\rm max}$(maximum electron Lorentz factor)= $2\cdot10^7$ 
   $\gamma_{\rm b}$(break electron Lorenz factor)=  $5\cdot10^4$ 
   the slopes of the electron distribution $n_1 = 2$ and 
   $n_2 = 5$ before and after the break energy, respectively,
   as well as 
  $n_{\rm e}$(normalization of the electron energy distribution) = $2\cdot10^4$\,cm$^{-3}$.
   The model is not intended for describing the radio data,
   which is assumed to origin from a larger emitting volume to avoid
   an intrinsic absorption.}
\end{figure}


As discussed in \citet{tavecchio98}, if the position and the luminosity
of the synchrotron and SSC peaks are known and an estimate of the
minimum variability timescale is available, it is possible to uniquely
constrain the model parameters. Unfortunately, we do not
have all the required information to accomplish this.  In particular, we 
have fixed the synchrotron peak by
requiring that it reproduces the optical flux and the historical X-ray
spectrum and we assume the SSC peak to be close to the
MAGIC threshold.  These choices minimize the required emitted
luminosity, since a lower SSC peak frequency would require a
higher SSC luminosity.

We present two models. The first (solid line), assuming an
electron distribution extending down to $\gamma _{\rm min}=1$, clearly
overpredicts the MeV-GeV flux
measured by EGRET. In the second case (dashed line), which fixes the
low energy limit at $\gamma _{\rm min}=3\cdot 10^3$ (leading to a
``narrowing'' of both the synchrotron and SSC bump, see
\citet{katarzynski}) the model is compatible with the reported EGRET
flux. It is evident that simultaneous \textit{GLAST}-MAGIC observations of this
source will provide important constraints on the model parameters.

In both cases, the energy output of the SSC component (reaching 
observed values of $L\sim
10^{46}$ erg/s) dominates over the synchrotron luminosity, implying a
relatively low magnetic field, $B=0.15$ G. In that case the source
would be strongly electron dominated, since the magnetic energy
density would be several orders of magnitude below that of the
relativistic electrons. A larger synchrotron flux (limited by the
non-detection by BAT and ASM) could alleviate the
problem. Simultaneous X-ray and VHE observations are mandatory to
further investigate this issue.  We also note the fit Doppler factor
($\delta=20$) is rather high and should be verified by Very Long
Baseline Interferometric observations. The fitted parameters are
similar to those derived for other TeV-emitting BL Lacs. We note
however that adopting models where the jet has a velocity structure
(e.g. models by \citet{GK,Ghi05}) would considerably reduce the
required Doppler factors.

1ES~1011+496 was previously observed with the HEGRA telescope array,
resulting in an upper limit of F(E$>$1 TeV)$\le 1.8\cdot 10^{-12}$
photons cm$^{-2}$s$^{-1}$ (Aharonian et al. 2004), which is well above
the detected flux we found. The source was also observed by MAGIC, as part of a
systematic scan of X-ray-bright HBLs, in March--April 2006.
Being in a lower optical state (the core flux was $\sim50\%$ lower than
that in March--May 2007), the observations showed a marginal signal with
$3.5\sigma$ significance corresponding to
an integral flux of F($>$180~GeV)=(1.26$\pm0.4)\cdot 10^{-11}$
photons cm$^{-2}$s$^{-1}$, i.e.\ $\sim 40\%$ \citep{hbl} lower than the detected
flux in March--May 2007 (see also the inset in Fig.~4). 
A similar trend was also found for BL~Lac \citep{bllac}, where the
observations during a lower optical state failed to detect VHE
$\gamma$-rays. This seems to indicate that there is a connection
between the optical high state and the higher flux of VHE $\gamma$-ray
emission at least in some sources. To further investigate this
possibility, follow-up
observations of the detected objects as well as further observations
of other AGNs during high optical states are required.


We thank the IAC for the excellent working conditions at the ORM. The
support of the German BMBF and MPG, the Italian INFN, the Spanish
CICYT, the ETH Research Grant TH~34/04~3 and the Polish MNiI Grant
1P03D01028 is gratefully acknowledged.  We thank A. Berdyugin for
performing the optical observations and A. L\"ahteenm\"aki
(Mets\"ahovi) for the radio observations.

\end{document}